\documentclass[pre,epsfig,address,twocolumn,showpacs]{revtex4}%
\usepackage[english]{babel}
\usepackage[latin1]{inputenc}
\usepackage[dvips]{graphicx}
\usepackage{amsmath}
\usepackage{amssymb}
\usepackage{epsfig}
\usepackage{array}
\usepackage{amsfonts}
\usepackage{graphicx}
\usepackage{subfigure}%
\begin{document}
\title{Manipulating energy and spin currents in nonequilibrium systems of interacting qubits}
\author{V. Popkov$^{1,3}$and R. Livi$^{1,2}$}
\affiliation{$^{1}$ Dipartimento di Fisica e Astronomia -- CSDC, Universit\'a degli Studi
di Firenze, via G. Sansone 1, 50019 Sesto Fiorentino, Italy }
\affiliation{$^{2}$ INFN Sezione di Firenze, via G. Sansone 1, 50019 Sesto Fiorentino,
Italy }
\affiliation{$^{3}$ Max Planck Institute for Complex Systems, N\"othnitzer Stra{ß}e 38,
01187 Dresden, Germany}
\date{\today }

\begin{abstract}
We consider generic interacting chain of qubits, which are coupled at the
edges to baths of fixed polarizations. We can determine the nonequilibrium
steady states, described by the fixed point of the Lindblad Master Equation.
Under rather general assumptions about local pumping and interactions,
symmetries of the reduced density matrix are revealed. The symmetries
drastically restrict the form of the steady density matrices in such a way
that an exponentially large subset of one--point and many--point correlation
functions are found to vanish. As an example we show how in a Heisenberg spin
chain a suitable choice of the baths can completely switch off either the spin
or the energy current, or both of them, despite the presence of large boundary gradients.

\end{abstract}

\maketitle

\section{ Introduction}

The impressive progress in experimental manipulation of nanowires and quantum
dots makes it possible to investigate quantum systems consisting of a few
quantum dots or quantum bits. On the other hand, manipulations/operations on a
single quantum bit (e.g., application of a quantum gate) are at the basis of
the functioning of any elementary block of a quantum computing device.
However, a theoretical understanding of microscopic quantum systems out of
equilibrium\ (e.g. under constant pumping or continuous measurement by a
quantum probe) is far from being complete, apart from simple cases like a
single two-level system or a quantum harmonic oscillator under external
pumping or in contact with reservoirs \cite{Petruccione,PlenioJumps}. On the
other hand, using the dissipative dynamics for the preparation of quantum
states with required properties is becoming a promising field of research
\cite{ZollerNature2008,ZollerPRA}. In this respect, the role of sufficiently
simple spatially extended systems, amenable to both analytic and numerical
investigations, becomes important.

Unlike a quantum system evolving coherently, whose evolution at any time
depends on the initial state, a quantum system with pumping tends to a steady
state, independently of the initial conditions. This allows to manipulate the
steady state of the system by a suitable choice of the pumping. In this paper
we first illustrate how to accomplish such a task for a general system of
interacting quantum spins. In particular we consider symmetries of the density
matrix that impose rigid constraints on the properties of the nonequilibrium
steady state, thus entailing, for instance, exact vanishing of cumulative
correlation functions, like certain components of the structure factor. Then,
we specialize our analysis to the one dimensional driven XXZ chain of quantum
spins in the presence of pumping applied at the edges. We show that, for
particular realizations of the pumping and irrespectively of the system size,
one can switch on and off the spin and/or the heat currents in the
nonequilibrium steady-state. Such an approach unveils new interesting
perspectives in the study of driven spin chain models. They have been mainly
investigated to understand under which conditions the spin and the energy
currents in the steady state exhibit anomalous or ballistic behavior
\cite{reviewBrenig07,zotos,KlumperLectNotes2004}. In this regard, we want to
point out that the results presented in this paper hold independently of the
anomalous or ballistic features of transport \cite{ZotosBallistic97} and even
in the absence of integrability.

In Section \ref{sec::Model} we present the general model of Lindblad master
equation and discuss how symmetries may affect its properties. Specific forms
of Lindblad operators acting on quantum spin models are introduced in Section
\ref{sec::Lindblad operators}. In particular, we devote special attention to
parity selection rules in Section \ref{sec::Parity symmetry selection rules}.
Section \ref{sec::1D driven XXZ model} deals with the special case of the
Lindblad dynamics for the one dimensional XXZ model of qubits. It has been
chosen to illustrate in detail how our findings apply to a simple model, that
has been the object of intense resent research. Specifically, in subsection
\ref{subsec::Parity symmetry and energy current} we discuss the kind of
symmetries emerging in this model for Lindblad operators of target type, while
subsection \ref{subsec::Spin and thermal conductance with boundary gradients}
is devoted to the analysis of spin and energy conductance in the presence of
gradients. Conclusions and perspectives are contained in Section
\ref{sec::Conclusions}.

\section{Lindblad Master Equation and its symmetries}

\label{sec::Model}

We consider the quantum Master equation in the Lindblad form
\cite{Petruccione,Wichterich07},
\begin{equation}
\frac{\partial\rho}{\partial t}=i\left[  \rho,H\right]  -\frac{1}{2}\sum
_{m}\left\{  \rho,L_{m}^{(p)\dagger}L_{m}^{(p)}\right\}  +\sum_{m,p}%
L_{m}^{(p)}\rho L_{m}^{(p)\dagger}, \label{LindbladMasterEquation}%
\end{equation}
where we have set $\hbar=1$; $\rho$ is the reduced density matrix, $H$ is the
Hamiltonian of the system and $L_{m}^{(p)}$ is the Lindblad operator. It is
easy to verify
%from (\ref{LindbladMasterEquation}),
that $\frac{\partial}{\partial t}Tr(\rho)=0,\rho^{+}=\rho,\frac{\partial
}{\partial t}Tr(\rho^{2})\neq0$. The first two relations are necessary for
interpreting $\rho$ as a density matrix, with $Tr(\rho)=1$, while the latter
implies that we deal with an open system: an initially pure state $\rho
=|\phi\rangle\langle\phi|$ will not remain pure in the course of time. The
Lindblad equation is the most general Markovian equation of motion for the
reduced density matrix, conserving positivity and trace, and having a
semigroup property \cite{Petruccione}.

The non-unitary part of the Lindblad Master equation (LME) makes the dynamics
irreversible. In the course of time, any initial state will relax to a
nonequilibrium steady state (NESS), described by the time-independent solution
of the Master equation. Our purpose is to reveal symmetries of the LME\ and
respective constraints on the NESS, which the symmetries impose. In the
simplest case, the constraints on NESS take the form of selection rules,
according to which a subset of the matrix elements of the reduced density
matrix has to vanish.

Let us denote the right-hand side of LME as $%
%TCIMACRO{\tciLaplace}%
%BeginExpansion
\mathcal{L}%
%EndExpansion
\lbrack\rho]$. $\ $If it is invariant under a unitary transformation $U$, i.e.
$%
%TCIMACRO{\tciLaplace}%
%BeginExpansion
\mathcal{L}%
%EndExpansion
\lbrack U\rho U^{\dagger} ]=U%
%TCIMACRO{\tciLaplace}%
%BeginExpansion
\mathcal{L}%
%EndExpansion
\lbrack\rho]U^{\dagger} $, it follows that $\tilde{\rho}(t)=U\rho
(t)U^{\dagger} $ is a new solution of LME. In general the solutions
$\tilde{\rho}(t)$ and $\rho(t)$ describe different time evolutions. If the
nonequilibrium steady state (NESS) of the system $\rho_{NESS}=\lim
_{t\rightarrow\infty}\rho(t)$ is unique \cite{EvansUniqueness}, then the
trajectories $\rho(t)$ and $\tilde{\rho}(t)$ eventually converge in time to
the same asymptotic solution, $\lim_{t\rightarrow\infty}\rho(t)=\lim
_{t\rightarrow\infty}\tilde{\rho}(t)=\rho_{NESS}$. Accordingly, the unique
steady state of the system has to be invariant under the transformation $U$,
\begin{equation}
\rho_{NESS}=U\rho_{NESS}U^{\dagger} . \label{SymmetryRho}%
\end{equation}
This implies also that the expectation value of any physical observable of
interest $\hat{f}$, measured in the steady state, $\langle\hat{f}\rangle\equiv
Tr\left(  \hat{f}\rho_{NESS}\right)  $, has to satisfy the relations
\begin{equation}
\langle\hat{f}\rangle=Tr\left(  \hat{f}U\rho_{NESS}U^{\dagger} \right)
=\langle U^{\dagger} \hat{f}U\rangle.
\end{equation}
.In particular, if $\hat{f}$ changes sign under the action of $U$,
\begin{equation}
U^{\dagger} \hat{f}U=-\hat{f}, \label{Ustar_f_U=-f}%
\end{equation}
it follows that
\begin{align}
\langle\hat{f}\rangle &  =Tr\left(  U^{\dagger} \hat{f}U\rho_{NESS}\right)
\nonumber\\
&  =-Tr\left(  \hat{f}\rho_{NESS}\right)  =-\langle\hat{f}\rangle
\,\,\rightarrow\,\,\langle\hat{f}\rangle=0\,\,.
\end{align}
On the contrary, if $\hat{f}$ is invariant under the action of $U$, i.e.
\begin{equation}
U^{\dagger} \hat{f}U=\hat{f}\,\,,
\end{equation}
no consequences for $\langle\hat{f}\rangle$ can be drawn,
\begin{equation}
\langle\hat{f}\rangle=Tr\left(  U^{\dagger} \hat{f}U\rho_{NESS}\right)
=Tr\left(  \hat{f}\rho_{NESS}\right)  \,\,.
\end{equation}

For instance, let us suppose that a two-level open quantum system, described
by a generic density matrix $\rho=\frac{1}{2}I+\frac{1}{2}\sum_{\alpha}%
\langle\sigma^{\alpha}\rangle\sigma^{\alpha}$ ($\sigma^{\alpha}$ being Pauli
matrices), is invariant under the transformation $U=\sigma^{z}$, and has a
unique steady state. Since $\sigma^{z}\sigma^{x}\sigma^{z}=-\sigma^{x}$,
$\sigma^{z}\sigma^{y}\sigma^{z}=-\sigma^{y}$ and $\sigma^{z}\sigma^{z}%
\sigma^{z}=\sigma^{z}$, from (\ref{Ustar_f_U=-f}) we can conclude that
$\langle\sigma^{x}\rangle=\langle\sigma^{y}\rangle=0$, while $\langle
\sigma^{z}\rangle$ is invariant under the action of $U$.

\section{ Lindblad operators}

\label{sec::Lindblad operators}

Here we focus on the most commonly used types of Lindblad operators: (i) those
targeting a given value of observables $\langle\sigma_{p}^{\alpha}\rangle$;
(ii) those responsible for dephasing, or decoherence, whose action favors the
evolution to a classical state; (iii) the non-local variants of the previous cases.

Lindblad operators targeting a given value of $z-$ spin projection
$\sigma_{\text{target}}^{z}$ at a chosen site $p$ have the form of
creation-annihilation operators,
\begin{align}
L_{1}^{(p)}  &  =\alpha(\sigma_{p}^{x}+i\sigma_{p}^{y})=\alpha\sigma_{p}%
^{+}\label{L_targetingZ_0}\\
L_{2}^{(p)}  &  =\beta(\sigma_{p}^{x}-i\sigma_{p}^{y})=\beta\sigma_{p}^{-}.
\label{L_targetingZ}%
\end{align}

The equations of motion for the expectation values of the operators
$\sigma_{p}^{x},\sigma_{p}^{y},\sigma_{p}^{z}$ \ read
\begin{align}
\frac{d\langle\sigma_{p}^{z}\rangle}{dt}  &  ={\mathcal{H}}(\sigma_{p}%
^{z})-\Gamma_{z}(\langle\sigma_{p}^{z}\rangle-\sigma_{\text{target}}%
^{z})\label{LeftBorderEquationOfMotion}\\
\frac{d\langle\sigma_{p}^{x}\rangle}{dt}  &  ={\mathcal{H}}(\sigma_{p}%
^{x})-\Gamma_{x}\langle\sigma_{p}^{x}\rangle\nonumber\\
\frac{d\langle\sigma_{p}^{y}\rangle}{dt}  &  ={\mathcal{H}}(\sigma_{p}%
^{y})-\Gamma_{y}\langle\sigma_{p}^{y}\rangle\,\, ,\nonumber
\end{align}
where we introduce the shorthand notation ${\mathcal{H}}(f_{p}) \equiv
-iTr\left(  f_{p}\left[  H,\rho\right]  \right)  $, and
\begin{align}
\Gamma_{z}  &  =4(\alpha^{2}+\beta^{2});\text{ \ }\sigma_{\text{target}}%
^{z}=\frac{\alpha^{2}-\beta^{2}}{\alpha^{2}+\beta^{2}}%
\label{BoundarySpinValues}\\
\Gamma_{x}  &  =\Gamma_{y}=\Gamma_{z}/2 \, .
\end{align}

If the coupling constant $\Gamma_{z}$ is sufficiently large with respect to
the norm of the Hamiltonian in equations (\ref{LeftBorderEquationOfMotion}),
then ${\mathcal{H}}$ can be neglected and the averages $\langle\sigma_{p}%
^{i}(t)\rangle$ converge, after some relaxation time of order $1/\Gamma_{\eta
},$ to their "targeted" values $\sigma_{\text{target}}^{x}=0,\sigma
_{\text{target}}^{y}=0,\sigma_{\text{target}}^{z}=\left(  \alpha^{2}-\beta
^{2}\right)  /\left(  \alpha^{2}+\beta^{2}\right)  $ as follows
\begin{equation}
\langle\sigma_{p}^{\eta}(t)\rangle=\sigma_{\text{target}}^{\eta}+\langle
\sigma_{p}^{\eta}(0)-\sigma_{\text{target}}^{\eta}\rangle e^{-\Gamma_{\eta}%
t}\,. \label{BoundaryRelaxation}%
\end{equation}

Lindblad operators of type (\ref{L_targetingZ}) naturally appear in the
problem of an atom interacting with a quantized radiation field
\cite{GardinerZoller2000}, in spin chains coupled to a bath of fixed
polarization, in electron paramagnetic resonance experiments and in studies of
quantum transport
\cite{Pros08,ZunkovichJStat2010ExactXY,BenentiPRB2009,ProsenExact2011,JesenkoZnidaric2011,Casati2009,ProsenNY2010,
Znidaric2011,ProsenZnidaric,ZnidaricJStat2010}.

Lindblad operators targeting a given value of $x-$spin projection or $y-$spin
projection are given by the appropriate cyclic rotation of the operators
(\ref{L_targetingZ}), i.e.
\begin{equation}
V_{1}=\alpha(\sigma_{p}^{y}+i\sigma_{p}^{z}),\text{ \ }V_{2}=\beta(\sigma
_{p}^{y}-i\sigma_{p}^{z})\text{ } \label{L_targetingX}%
\end{equation}
with target $\sigma_{\text{target}}^{x}=\left(  \alpha^{2}-\beta^{2}\right)
/\left(  \alpha^{2}+\beta^{2}\right)  $ at site $p$ and%
\begin{equation}
W_{1}=u(\sigma_{p}^{z}+i\sigma_{p}^{x}),\text{ \ }W_{2}=v(\sigma_{p}%
^{z}-i\sigma_{p}^{x})\text{ } \label{L_targetingY}%
\end{equation}
with target $\sigma_{\text{target}}^{y}=\left(  u^{2}-v^{2}\right)  /\left(
u^{2}+v^{2}\right)  $ at site $p$. The relaxation to the targeted values can
be described in complete analogy with (\ref{BoundarySpinValues}%
)-(\ref{BoundaryRelaxation}), see \cite{Lindblad2011}.

Dephasing Lindblad operators have the form
\begin{equation}
L_{deph}^{(p)}=\sqrt{\gamma_{p}}\sigma_{p}^{z} \label{L_dephasing}%
\end{equation}
and describe the presence of a dephasing noise in the dynamics, which causes
decoherence (i.e., vanishing of non-diagonal density matrix elements) with
rate $\gamma_{p}$ at site $p$. The local equations of motion, analogous to
(\ref{LeftBorderEquationOfMotion}), have the simple form
\begin{equation}
\frac{d\langle\sigma_{p}^{x,y}\rangle}{dt}=-2\gamma_{p}\langle\sigma_{p}%
^{x,y}\rangle\,\,\,,\,\,\,\frac{d\langle\sigma_{p}^{z}\rangle}{dt}=0\,.
\end{equation}
The coupling to a heat bath is often modelled by the application of the
dephasing Lindblad operators (\ref{L_dephasing}) at all sites of the system.

Lindblad operators, acting on more than one site, introduce incoherent
non-local processes. For instance the Lindblad operator
\begin{equation}
L_{hopp}^{(p,q)}=\sqrt{\gamma}\sigma_{p}^{+}\sigma_{q}^{-} \label{L_hoppings}%
\end{equation}
describes incoherent spin flips between sites $p$ and $q$
\cite{EislerBallistic}.

\section{Parity symmetry selection rules for Lindblad dynamics}

\label{sec::Parity symmetry selection rules}

We consider a generic open system of qubits, with internal pair interactions,
described by the Lindblad Master equation (\ref{LindbladMasterEquation}), with
the Hamiltonian%

\begin{align}
\label{Hgen}H  &  =\sum_{k,m=1}^{N}J_{X}(k,m)\sigma_{k}^{x}\sigma_{m}%
^{x}+J_{Y}(k,m)\sigma_{k}^{y}\sigma_{m}^{y}\\
&  +J_{Z}(k,m)\sigma_{k}^{z}\sigma_{m}^{z}+\sum_{k=1}^{N}h_{k}\sigma_{k}%
^{z}\,\,, \nonumber\label{HamiltonianHeisenberg}%
\end{align}
where $J_{\alpha}(k,m)$ are couplings between qubits $k,m$, and $h_{k}$ are
local magnetic fields. We do not impose any restriction on the space dimension
or on the geometry, but we just assume the connectivity of the graph. The
system can be subject to external pumping and/or external noise, modelled by
the $z$-polarization targeting operators (\ref{L_targetingZ}), dephasing
Lindblad operators (\ref{L_dephasing}) and incoherent hoppings
(\ref{L_hoppings}), with all of these operators acting on an arbitrary subset
of sites. Then, if the steady state is unique, the transformation $\Omega
_{z}=(\sigma^{z})^{\otimes_{N}}=\sigma^{z}\otimes\sigma^{z}\otimes
...\otimes\sigma^{z}$ , $\Omega_{z}^{-1}=\Omega_{z}$ identifies a symmetry of
the Master equation, thus yielding the relation
\begin{equation}
\rho_{NESS}=\Omega_{z}\rho_{NESS}\Omega_{z}. \label{ParityConservation}%
\end{equation}
Indeed, Hamiltonian (\ref{Hgen}) as well as dephasing Lindblad operators
(\ref{L_dephasing}) and incoherent hoppings (\ref{L_hoppings}) are invariant
under $\Omega_{z}$, while $z-$polarization targeting operators
(\ref{L_targetingZ}) change sign under its action. Since the Lindblad part of
the evolution equation is quadratic in $L_{m}$, LME is invariant under
$\Omega_{z}$. The symmetry (\ref{ParityConservation}) is known in spin models
as parity symmetry (P--symmetry). In the special case where the total set of
Lindblad operators contains either only dephasing Lindblad operators
(\ref{L_dephasing}) or only z-polarization targeting Lindblad operators
(\ref{L_targetingZ}), LME has one further symmetry, the PT- symmetry, which
has interesting consequences on the full spectrum of the Lindblad
superoperator \cite{ProsenSymmetries}. In our general setting, where the
dephasing and the polarization targeting Lindblad operators are mixed, the
quantum Lindblad dynamics is not PT-invariant. Another remark concerns our
crucial assumption of a uniqueness of the steady state. The existence and
uniqueness of the steady state is guaranteed by the completeness of the
algebra, generated by the set of operators $\{H,L_{m},L_{m}^{\dagger} \}$
under multiplication and addition \cite{EvansUniqueness}, and it is verified
straightforwardly as in \cite{ProsenUniqueness}, for any choice of a set of
the Lindblad operators $\{L_{m}\}$, provided the set contains at least one
polarization targeting operator (\ref{L_targetingZ_0}) or (\ref{L_targetingZ}).

The P--symmetry yields severe limitations on the nonequilibrium steady state
$\rho_{NESS}$. We indicate with $\rho_{j_{1}j_{2}...j_{N}}^{i_{1}i_{2}%
...i_{N}}$ the matrix element of $\rho_{NESS}$ in the natural basis, labelled
by indexes $i_{1},i_{2},...j_{N}$ which take values $-1,1$. Let us calculate
how the matrix element $\rho_{j_{1}j_{2}...j_{N}}^{i_{1}i_{2}...i_{N}}$
changes under the action of the P--symmetry operator $\Omega_{z}$. One
obtains
\begin{equation}
\left(  (\sigma^{z})^{\otimes_{N}}\rho(\sigma^{z})^{\otimes_{N}}\right)
_{j_{1}j_{2}...j_{N}}^{i_{1}i_{2}...i_{N}}=\rho_{j_{1}j_{2}...j_{N}}%
^{i_{1}i_{2}...i_{N}}\times\prod\limits_{m=1}^{N}i_{m}j_{m}.
\end{equation}
The factor $K=\prod\limits_{m=1}^{N}i_{m}j_{m}$ may only take values $1$ and
$-1$. If $K=1$, the P--symmetry does not yield any constraint on $\rho$.
Conversely, if $K=-1$, from (\ref{ParityConservation}) it follows that the
corresponding matrix element vanishes,%
\begin{equation}
\rho_{j_{1}j_{2}...j_{N}}^{i_{1}i_{2}...i_{N}}\equiv0\,\,\,\,\,\text{ if
}\,\,\,\,\,\prod\limits_{m=1}^{N}i_{m}j_{m}=-1\text{.}
\label{ParitySelectionRule}%
\end{equation}
We call this condition a parity selection rule (PSR). If
(\ref{ParitySelectionRule}) holds, simple analysis shows that each row and
each column of $\rho$ contains $2^{N-1}$ zero entries. For instance, the first
row of the density matrix for 3 sites $\rho_{\alpha\beta\gamma}^{111}$
contains four null elements: $\rho_{1\text{ }1\text{ }-1}^{1\text{ }1\text{
}1}=\rho_{-1\text{ }1\text{ }1}^{1\text{ }1\text{ }1}=\rho_{1\text{ }-1\text{
}1}^{1\text{ }1\text{ }1}=\rho_{-1\text{ }-1\text{ }-1}^{1\text{ }1\text{ }%
1}=0$. More generally, it can be easily realized that by a suitable
reshuffling of rows/columns the $2^{N}\times2^{N}$ density matrix satisfying
(\ref{ParitySelectionRule}) can be brought into a block-diagonal form, with
two blocks of equal size $2^{N-1}\times2^{N-1}$.

An important feature of PSR is that any subsystem made of $n$ qubits,
described by the reduced density matrix $\rho_{(n)}=Tr_{N-n}\rho$, keeps the
same symmetry, as it can be easily verified,
\begin{equation}
\rho_{(n)j_{1}j_{2}...j_{n}}^{\text{ \ \ \ \ }i_{1}i_{2}...i_{n}}\equiv0\text{
for }\prod\limits_{m=1}^{n}i_{m}j_{m}=-1.
\end{equation}
For $n=2$ the explicit form of the generic density matrix is
\begin{equation}
\rho=%
\begin{pmatrix}
a & 0 & 0 & b\\
0 & c & d & 0\\
0 & d^{\ast} & c_{1} & 0\\
b^{\ast} & 0 & 0 & a_{1}%
\end{pmatrix}
, \label{X_states}%
\end{equation}

States of the form (\ref{X_states}) are well known in information theory as
$X$-states and they are subject of intensive investigation, (see e.g.
\cite{DiscordXstates}).

Finally, we want to point out that, in terms of physical observables, PSR
(\ref{ParitySelectionRule}) entails vanishing of a set of experimentally
measurable quantities, like many-point correlation functions $\langle
\sigma_{m_{1}}^{\alpha}\sigma_{m_{2}}^{\beta}...\sigma_{m_{k}}^{\gamma}%
\rangle$, and structure factors $S^{\alpha\beta}(k,\Delta)=\sum_{n<m}%
e^{ik(m-n)}\langle\sigma_{n}^{\alpha}\sigma_{n+1}^{\beta}\rangle$, namely
\begin{equation}
\langle\sigma_{n}^{x}\rangle=\langle\sigma_{n}^{y}\rangle=0
\label{OnePointCorrVanish}%
\end{equation}

\begin{equation}
\langle\sigma_{n}^{y}\sigma_{m}^{z}\rangle=\langle\sigma_{n}^{x}\sigma_{m}%
^{z}\rangle=0\text{ } \label{TwoPointCorrVanish}%
\end{equation}%
\begin{equation}
\langle\sigma_{n}^{y}\sigma_{m_{1}}^{z}\sigma_{m_{2}}^{z}...\sigma_{m_{k}}%
^{z}\rangle=\langle\sigma_{n}^{x}\sigma_{m_{1}}^{z}\sigma_{m_{2}}^{z}%
...\sigma_{m_{k}}^{z}\rangle=0\text{ }%
\end{equation}%
\begin{align}
\langle\sigma_{n_{1}}^{x}\sigma_{n_{2}}^{x}\sigma_{m}^{y}\rangle &
=\langle\sigma_{n}^{x}\sigma_{m_{1}}^{y}\sigma_{m_{2}}^{y}\rangle=0,\text{ }\\
&  ...\nonumber
\end{align}%
\begin{equation}
S^{xz}(k)=S^{yz}(k)=S^{zx}(k)=S^{zy}(k)=0. \label{StructureFactorsVanish}%
\end{equation}
In Refs \cite{Krammer09} and \cite{CramerPlenio11} the structure factors
$S^{\alpha\beta}(k,\Delta)$ were proposed as entanglement witnesses: they are
measurable quantities in neutron scattering experiments.

\section{$1D$ driven $XXZ$ model}

\label{sec::1D driven XXZ model}

\subsection{Parity symmetry and energy current}

\label{subsec::Parity symmetry and energy current}

The general properties described in the previous section can be specialized to
the study of further symmetries emerging in one-dimensional driven spin chain
models with pumping acting at the edges. In these cases Lindblad operators
create effective boundary gradients. A commonly studied setup (see
\cite{Pros08, ZunkovichJStat2010ExactXY,BenentiPRB2009,
ProsenExact2011,JesenkoZnidaric2011,Casati2009,ProsenNY2010,Znidaric2011,ProsenZnidaric}
) is the $XXZ$ spin chain, whose Hamiltonian reads
\begin{equation}
H=\sum_{k=1}^{N-1}\sigma_{k}^{x}\sigma_{k+1}^{x}+\sigma_{k}^{y}\sigma
_{k+1}^{y}+\Delta\sigma_{k}^{z}\sigma_{k+1}^{z}. \label{hxxz}%
\end{equation}
There are four Lindblad operators, $L_{1,2}=\sqrt{\Gamma(1+\mu)}\sigma
_{1}^{\pm}$ and $L_{3,4}=$ $\sqrt{\Gamma(1-\mu)}\sigma_{N}^{\pm}$,
parametrized by $-1\leq\mu\leq1$ and $\Gamma>0$, that target at the chain
edges spin configurations with equal amplitude and opposite sign,
$\langle\sigma_{1}^{z}\rangle\rightarrow\mu$, $\langle\sigma_{N}^{z}%
\rangle\rightarrow-\mu$, see Eq. (\ref{BoundarySpinValues}). In addition to
the P--symmetry (\ref{ParityConservation}) \cite{RemarkPsymmetry}, discussed
in the previous section and the PT--symmetry discussed in
\cite{ProsenSymmetries}, in this case there exist an additional symmetry
\begin{equation}
\rho_{NESS}=\Omega_{x}R\rho_{NESS}R\Omega_{x} \label{OmegaXsymmetry}%
\end{equation}
where $R(A\otimes B\otimes...\otimes C)=(C\otimes....\otimes B\otimes A)R$ is
a left-right reflection and $\Omega_{x}=(\sigma^{x})^{\otimes_{N}}$
\cite{RemarkOmegaYsymmetry}. We shall see that the symmetry
(\ref{OmegaXsymmetry}) imposes restrictions on transport properties in the
driven $XXZ$ chain. The transport properties are governed by the spin and the
energy current operators, $\hat{\jmath}_{n,m}$ and $\hat{J}_{n}^{E}$, which
are defined by the lattice continuity equations $\frac{d}{dt}\sigma_{n}%
^{z}=\hat{\jmath}_{n-1,n}-\hat{\jmath}_{n,n+1}$, \ $\frac{d}{dt}h_{n,n+1}%
=\hat{J}_{n}^{E}-\hat{J}_{n+1}^{E}$, where
\begin{equation}
\hat{\jmath}_{n,m}=2(\sigma_{n}^{x}\sigma_{m}^{y}-\sigma_{n}^{y}\sigma_{m}%
^{x}) \label{Jz}%
\end{equation}
and
\begin{equation}
\hat{J}_{n}^{E}=-\sigma_{n}^{z}\hat{\jmath}_{n-1,n+1}+\Delta(\hat{\jmath
}_{n-1,n}\sigma_{n+1}^{z}+\sigma_{n-1}^{z}\hat{\jmath}_{n,n+1}). \label{Q}%
\end{equation}
It can be easily checked that the energy current operator $\hat{J}_{n}^{E}$
changes sign under the action of (\ref{OmegaXsymmetry}), $\Omega_{x}R\hat
{J}_{n}^{E}R\Omega_{x}=-\hat{J}_{n}^{E}$, thus implying that in the steady
state $\langle\hat{J}_{n}^{E}\rangle=0$ for any system size. On the other
hand, the magnetization current (\ref{Jz}) is invariant under the above
transformation, $\ \Omega_{x}Rj_{n}R\Omega_{x}=j_{n}$ and therefore it is
allowed to flow. So magnetization current can flow and the energy current is
suppressed completely, despite the presence of boundary gradients. Another
simple consequence of the $\Omega_{x}R$ symmetry is obtained by applying it to
the total $z$-magnetization operator $S^{z}=\sum_{n}\sigma_{n}^{z}$: it
changes sign under the action $\Omega_{x}RS^{z}R$ $\Omega_{x}=-S^{z}$,
entailing that the NESS belongs to zero total magnetization sector,
reproducing a known result, see, e.g., \cite{ProsenSymmetries}.

\begin{figure}[ptb]
\centerline{
\includegraphics[width=0.5\textwidth]{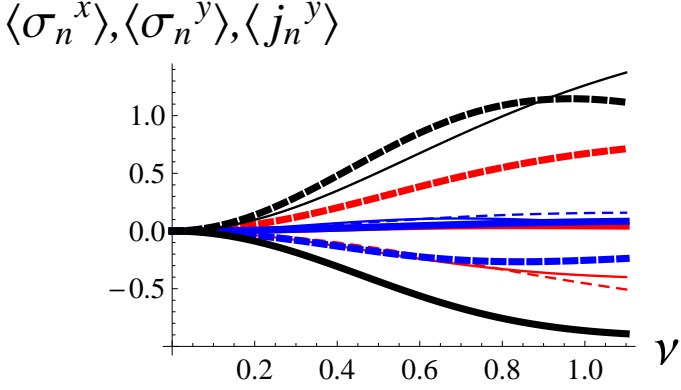}}\caption{Some
observables characterizing the NESS for an open $XXZ$ chain with Lindblad
operators $L_{1}=2\sigma_{1}^{+}$, $L_{2}=L_{3}=0$,$L_{4}=\sqrt{2}\sigma
_{N}^{-}$, $V=\nu(\sigma_{N}^{y}-i\sigma_{N}^{z})$, versus the amplitude $\nu
$. Red, blue and black lines correspond to one-point correlations
$\langle\sigma_{n}^{x}\rangle$ , $\langle\sigma_{n}^{y}\rangle$ and
correlations $\langle j_{n}^{y}\rangle=2$ $\langle\sigma_{n}^{z}\sigma
_{n+1}^{y}-\sigma_{n}^{y}\sigma_{n+1}^{z}\rangle$, respectively. Blue and red
color: thick,thin,thick dashed, thin dashed lines correspond to $n=1,2,3,4$
respectively. Black color: thick,thin, and thick dashed lines stand for
$\langle j_{n}^{y}\rangle$ at $n=1,2,3$. (note that $j_{n}^{y}$ is not
conserved locally and is therefore depends on $n$). For $\nu=0$ the symmetry
(\ref{ParityConservation}) is restored and all above observables vanish due to
(\ref{ParitySelectionRule}). The adopted parameter values are $N=4,J_{X}%
=J_{Y}=1,J_{Z}=-1.3$. }%
\label{Fig_NuCoeff}%
\end{figure}

To validate our results, we have integrated numerically equation
(\ref{LindbladMasterEquation}) that contains, in addition to the operators
$L_{1}$-- $L_{4}$, also the operator $V=\nu(\sigma_{N}^{y}-i\sigma_{N}^{z})$,
acting at site $i=N$. For zero amplitude $\nu=0$, the model possesses the
P--symmetry (\ref{ParityConservation}) \cite{RemarkPsymmetry}, while the
$\Omega_{x}R$ symmetry is broken because of non-symmetricity of left-right
boundary amplitudes, see caption of Fig.\ref{Fig_NuCoeff}. If $\nu\neq0$, also
the P--symmetry (\ref{ParityConservation}) is broken. In Fig.
\ref{Fig_NuCoeff} we plot various one- and two-point correlations as a
function of the amplitude $\nu$ of the P-symmetry breaking operator $V$: as
expected they are found to vanish only for $\nu=0$.
%see (\ref{OnePointCorrVanish}),(\ref{TwoPointCorrVanish}).\includegraphics[width=0.5\textwidth]

A different choice of the Lindblad operators, namely $L_{1,2}=\sqrt
{\Gamma(1+\mu)}(\sigma_{1}^{y}\pm i\sigma_{1}^{z})$ and $L_{3,4}=$
$\sqrt{\Gamma(1-\mu)}(\sigma_{N}^{z}\pm i\sigma_{N}^{x})$ amounts to set a
boundary twisting gradient in the $XY$-plane: the P--symmetry
(\ref{ParityConservation}) is violated, but other symmetries appear,
predicting a phenomenon of a sign alternation of the magnetization current
with the system size (see \cite{Lindblad2011}, \cite{PopkovXYtwist}). For
specific solvable cases, the full NESS of a $XXZ$ spin chain (\ref{hxxz}) with
Lindblad driving at the edges can be obtained analytically, see
\cite{ProsenExact2011},\cite{MPA2013}.

\subsection{Spin and thermal conductance with boundary gradients}

\label{subsec::Spin and thermal conductance with boundary gradients}

There is a large interest in studying the conductance in low-dimensional
materials, due to the rich and often counterintuitive features they exhibit.
In this section we discuss in full generality how one can switch on and off
the magnetization and the energy currents by a suitable choice of boundary
reservoirs , i.e. Lindblad operators, acting on the $XXZ$ spin chain
(\ref{hxxz}).

Let us couple the $XXZ$ chain at the boundaries to baths of constant (but
different) magnetizations, so that the time evolution of the state becomes
dissipative and is described by LME%

\begin{equation}
\frac{\partial\rho}{\partial t}=-i\left[  H,\rho\right]  +\Gamma(%
%TCIMACRO{\tciLaplace}%
%BeginExpansion
\mathcal{L}%
%EndExpansion
_{L}[\rho]+%
%TCIMACRO{\tciLaplace}%
%BeginExpansion
\mathcal{L}%
%EndExpansion
_{R}[\rho]), \label{LME}%
\end{equation}
where $H$ is the $XXZ$ Hamiltonian of the open $XXZ$ chain with anisotropy
$\Delta$ (see Eq. (\ref{hxxz})) .
%\begin{equation}
%H=J\sum_{k=1}^{N-1}\sigma_{k}^{x}\sigma_{k+1}^{x}+\sigma_{k}^{y}\sigma
%_{k+1}^{y}+\Delta\sigma_{k}^{z}\sigma_{k+1}^{z}, \label{Hamiltonian}%
%\end{equation}
$%
%TCIMACRO{\tciLaplace}%
%BeginExpansion
\mathcal{L}%
%EndExpansion
_{L}[\rho]$ and $%
%TCIMACRO{\tciLaplace}%
%BeginExpansion
\mathcal{L}%
%EndExpansion
_{R}[\rho]$ are Lindblad dissipators $%
%TCIMACRO{\tciLaplace}%
%BeginExpansion
\mathcal{L}%
%EndExpansion
\lbrack\rho]=\sum_{k}L_{k}\rho L_{k}^{\dagger} -\frac{1}{2}\{\rho
,L_{k}^{\dagger} L_{k}\}$ acting on the leftmost ($k=1$) and on the rightmost
($k=N$) boundary spins, while $\Gamma$ denotes the interaction rate with the
dissipators. By choosing different parameter values of the boundary Lindblad
dissipators $%
%TCIMACRO{\tciLaplace}%
%BeginExpansion
\mathcal{L}%
%EndExpansion
_{L}[\rho]$ and $%
%TCIMACRO{\tciLaplace}%
%BeginExpansion
\mathcal{L}%
%EndExpansion
_{R}[\rho]$, spin gradients can be introduced. In particular, the Lindblad
dissipators target spin polarizations at site $1$ and at site $N$, described
by the one-site density matrices $\rho_{L}$ and $\rho_{R}$ satisfying
$\mathcal{L}_{L}[\rho_{L}]=0$, $\mathcal{L}_{R}[\rho_{R}]=0$, respectively.
For sufficiently large values of $\Gamma$ the reduced density matrix of the
system $\rho(t)$ evolves in time towards a nonequilibrium steady state density
matrix, $\rho_{NESS}$, such that $Tr_{2,3,...N}\rho_{NESS}\rightarrow\rho_{L}$
and $Tr_{1,2,...N-1}\rho_{NESS}\rightarrow\rho_{R}$.

\begin{figure}[ptb]
\centerline{
\includegraphics[width=90 mm,height=70 mm]{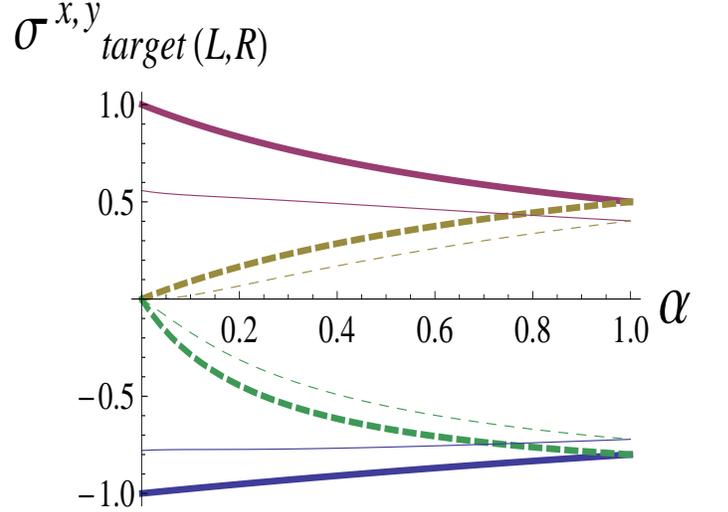}}\caption{Targeted
spin projections at the left and at the right boundary $\sigma
_{\text{target(L)}}^{x},\sigma_{\text{target(R)}}^{x}$(upper and lower bold
line, respectively), $\sigma_{\text{target(L)}}^{y},\sigma_{\text{target(R)}%
}^{y}$ (lower and upper dashed bold line, respectively) versus $\alpha$, from
(\ref{TargetedLeft}) and (\ref{TargetedRight}), with $A=2$. Thin full and
dashed lines mark the actual values of boundary magnetizations from numerical
LME solution, for the choice of parameters $A=2,N=5,\Delta=1,\Gamma=0.5$., see
also Fig.\ref{Fig_j_JE}. }%
\label{Fig_sxsyALPHA}%
\end{figure}

Let us choose the following Lindblad operators: for the left boundary, $%
%TCIMACRO{\tciLaplace}%
%BeginExpansion
\mathcal{L}%
%EndExpansion
_{L}$ dissipator contains operators
\begin{align}
L_{1}  &  =\sqrt{A}(\sigma_{1}^{y}-i\sigma_{1}^{z}),\\
L_{2}  &  =\sqrt{\alpha}(\sigma_{1}^{z}+i\sigma_{1}^{x}),
\end{align}
and for the right boundary, $%
%TCIMACRO{\tciLaplace}%
%BeginExpansion
\mathcal{L}%
%EndExpansion
_{R}$ contains%
\begin{align}
L_{3}  &  =(\sigma_{N}^{y}+i\sigma_{N}^{z}),\\
L_{4}  &  =\sqrt{A\alpha}(\sigma_{N}^{z}-i\sigma_{N}^{x}).
\end{align}
We assume also that $A\neq1$ and $0\leq\alpha\leq1$. It is straightforward to
check that $%
%TCIMACRO{\tciLaplace}%
%BeginExpansion
\mathcal{L}%
%EndExpansion
_{L}[\rho_{L}]=0$ where $\rho_{L}=\frac{I}{2}+\frac{1}{2}\sum_{\beta}%
\sigma_{\text{target(L)}}^{\beta}\sigma^{\beta}$ with
\begin{equation}
\sigma_{\text{target(L)}}^{x}=-\frac{2A}{2A+\alpha};\text{ }\sigma
_{\text{target(L)}}^{y}=\frac{2\alpha}{A+2\alpha};\text{ }\sigma
_{\text{target(L)}}^{z}=0. \label{TargetedLeft}%
\end{equation}
The entries of the set $\sigma_{\text{target(L)}}^{\beta}$ are targeted spin
components at the left boundary. At the right boundary, the targeted spin
components are
\begin{equation}
\sigma_{\text{target(R)}}^{x}=\frac{2}{2+\alpha A};\text{ }\sigma
_{\text{target(R)}}^{y}=-\frac{2\alpha A}{1+2\alpha A};\text{ }\sigma
_{\text{target(R)}}^{z}=0. \label{TargetedRight}%
\end{equation}
Graphs of the targeted spin components at the left and at the right boundaries
for $A=2$ are shown in Fig.\ref{Fig_sxsyALPHA}.

Due to the Heisenberg exchange interaction, one might expect that the presence
of a spin gradient yields non-vanishing spin and heat currents, given by the
Fourier law
\begin{align}
j^{\gamma}  &  =\chi_{\gamma\beta}\frac{\Delta s^{\beta}}{\Delta l}\\
J^{E}  &  =\chi_{\beta}^{E}\frac{\Delta s^{\beta}}{\Delta l},
\end{align}
where $\Delta s^{\beta}/\Delta l=\langle\sigma_{N}^{\beta}-\sigma_{1}^{\beta
}\rangle/(N-1)$ is the actual boundary gradient, $\chi_{\beta}^{E}$ and
$\chi_{\gamma\beta}$ are transport coefficients, $\gamma,\beta=x,y,z$ and
summation over repeated indexes is assumed. Note, that, due to quantum
fluctuations, the actual average boundary magnetizations are only approximated
by the respective targeted values, but do not coincide with them,
$\sigma_{\text{target(L)}}^{\beta}\neq\langle\sigma_{1}^{\beta}\rangle$
(compare the bold and thin lines in Fig. \ref{Fig_sxsyALPHA}), unless the rate
$\Gamma$ becomes large. The overall qualitative behaviour of the actual $x$-
and $y$- boundary gradients, at least for not very small $\Gamma$, is close to
the targeted one, and yields applied gradients for all values of $0\leq
\alpha\leq1$ , see also Fig. \ref{Fig_sxsyALPHA}. However, we find that for
$\alpha=0$ the steady spin current is identically zero, $\langle j\rangle=0$ ,
while $\langle J^{E}\rangle\neq0$. On the other hand, for $\alpha=1$ we obtain
the opposite scenario, i.e. $\langle J^{E}\rangle=0$ and $\langle j\rangle
\neq0$. In fact, for $\alpha=0$ the stationary solution of the Lindblad
equation $\rho_{NESS}$ is invariant under the following transformation,
\begin{equation}
\rho_{NESS}=\Omega_{x}\rho_{NESS}\Omega_{x}\,, \label{SymmetryAlpha0}%
\end{equation}
where $\Omega_{x}=(\sigma^{x})^{\otimes_{N}}$. Analogously, for $\alpha=1$,
$\rho_{NESS}$ is invariant under the transformation
\begin{equation}
\rho_{NESS}=\Omega_{x}U_{rot}R\rho_{NESS}RU_{rot}^{\dagger} \Omega_{x}
\label{SymmetryAlpha1}%
\end{equation}
where $R$ is again the left-right reflection, $(A\otimes B\otimes...\otimes
C)=(C\otimes....\otimes B\otimes A)R$, and the diagonal matrix $U_{rot}%
=diag(1,i)^{\otimes_{N}}$ is a rotation in $XY$ plane: $U_{rot}\sigma_{n}%
^{x}U_{rot}^{\dagger} =$ $\sigma_{n}^{y}$, $U_{rot}\sigma_{n}^{y}%
U_{rot}^{\dagger} =-\sigma_{n}^{x}$. The Hamiltonian part of LME, $-i\left[
H,\rho\right]  $, is also invariant under both transformations, while for the
Lindblad part the symmetries are satisfied due to the specific forms of $%
%TCIMACRO{\tciLaplace}%
%BeginExpansion
\mathcal{L}%
%EndExpansion
_{L}[\rho]$ and $%
%TCIMACRO{\tciLaplace}%
%BeginExpansion
\mathcal{L}%
%EndExpansion
_{R}[\rho]$ for $\alpha=0$ and $\alpha=1$.

\textbf{Case $\alpha=0$}. Making use of the symmetry (\ref{SymmetryAlpha0})
and of the properties of the Pauli matrices, we obtain the following
expressions for the magnetization and for the energy currents (in what follows
we use the shorthand notations $j$ and $J^{E}$ for these quantities),
\begin{equation}
j=Tr(\rho_{NESS}\hat{\jmath})=-Tr(\Omega_{x}\rho_{NESS}\Omega_{x}\hat{\jmath
})=-j
\end{equation}%
\begin{equation}
J^{E}=Tr(\rho_{NESS}\hat{J}^{E})=Tr(\Omega_{x}\rho_{NESS}\Omega_{x}\hat{J}%
^{E})=J^{E}%
\end{equation}
The first one of these relations implies $j=0$, while no restrictions are
imposed for $J^{E}$.

\textbf{Case $\alpha=1$}. We find the opposite situation: the energy current
$J^{E}$ under the transformation (\ref{SymmetryAlpha1}) changes sign, while no
restrictions are imposed for the magnetization current $j$. We conclude that
$J^{E}=0$.

\textbf{Case} $0<\alpha<1$. For any intermediate value of $\alpha$, neither
(\ref{SymmetryAlpha0}) nor (\ref{SymmetryAlpha1}) are satisfied. Consequently,
both magnetization and energy currents are allowed to flow.

%%%%/////////////// Two Figures //////////////////////////////
\begin{figure}[ptbh]
\begin{center}
\subfigure[\label{fig:a}]
{\includegraphics[width=0.5\textwidth]{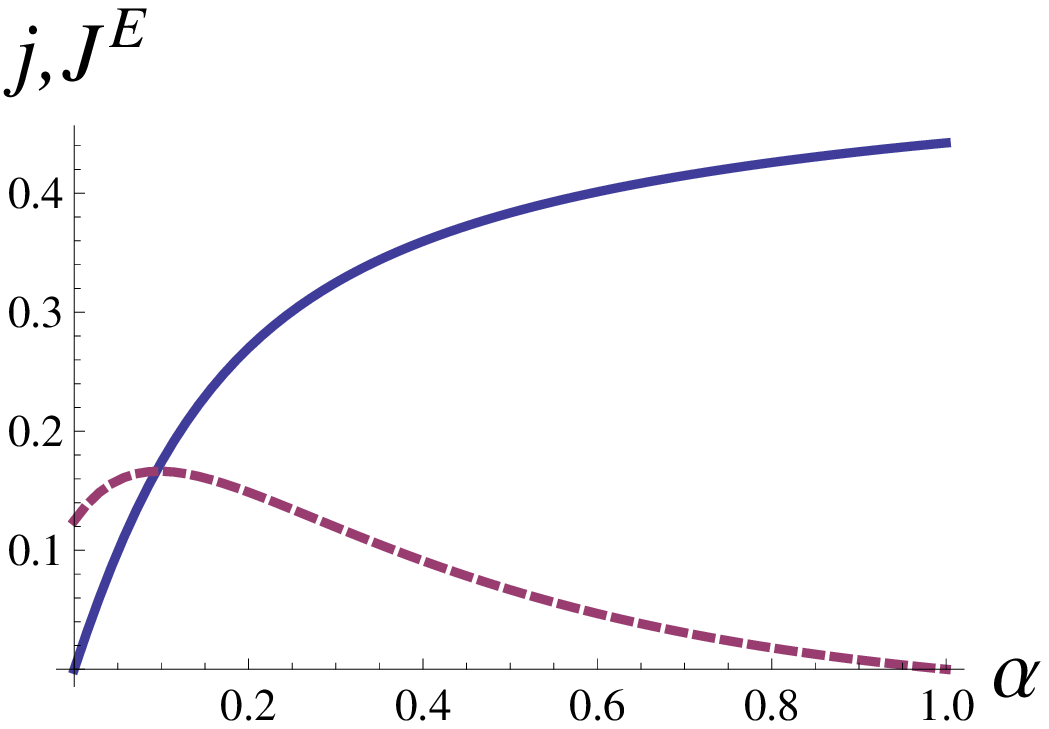}} \qquad
\subfigure[\label{fig:b}]{\includegraphics[width=0.5\textwidth]{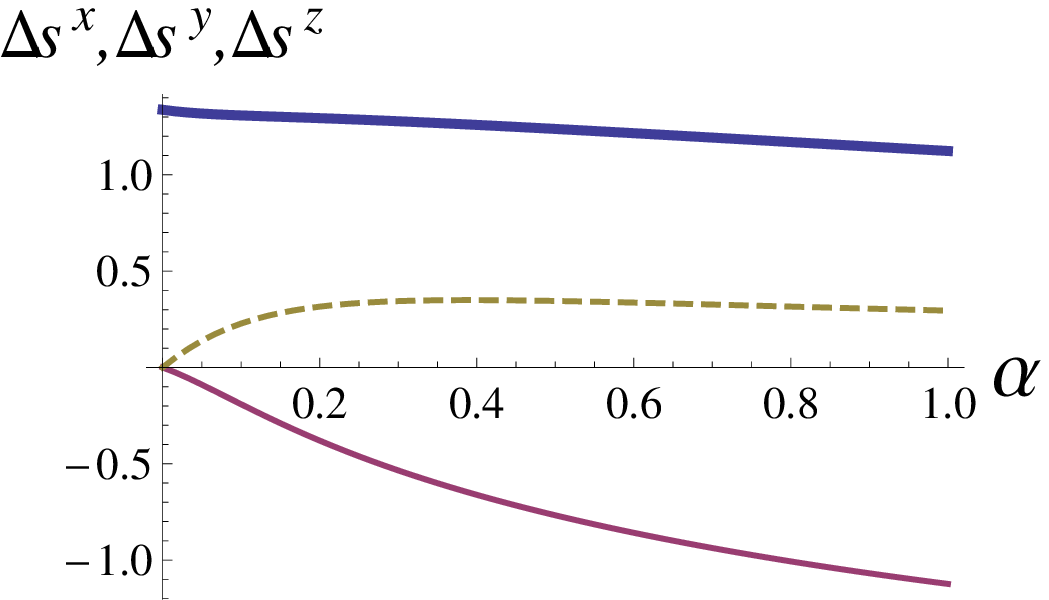}}
\end{center}
\caption{Panel(a): Steady magnetization current (line) and energy current
(dashed line) as a function of $\alpha$, from the numerical solution of the
Lindblad equation (\ref{LME}). Panel (b): Boundary $x,y,z$ gradients (thick,
thin and dashed lines respectively) $\langle\sigma_{N}^{x}-\sigma_{1}%
^{x}\rangle,\langle\sigma_{N}^{y}-\sigma_{1}^{y}\rangle,\langle\sigma_{N}%
^{z}-\sigma_{1}^{z}\rangle,$ versus $\alpha$. Note that the targeted values of
the boundary gradients fiven by difference of (\ref{TargetedRight}%
),(\ref{TargetedLeft}). are different, see also Fig. \ref{Fig_sxsyALPHA}, and
in particular the targeted value of $\ z$-gradient is $0$. The adopted
parameter values are $A=2,N=5,\Delta=1,\Gamma=0.5$.}%
\label{Fig_j_JE}%
\end{figure}

In order to check our findings, we have obtained numerical solutions of LME
(\ref{LME}) for small sizes $N$ and different values of the parameter
$\Delta\neq0$. In all cases we find complete agreement with theoretical
predictions. A typical case is illustrated in Fig.\ref{Fig_j_JE}. For $A=1$,
in addition we find that both $J_{\alpha=0,A=1}^{E}=0$ and $j_{\alpha
=0,A=1}=0$, while only $j=0$ is predicted by the symmetry
(\ref{SymmetryAlpha0}). Looking for an explanation, we readily find another
symmetry of (\ref{LME}), valid for $A=1$ and $\alpha=0$:
\begin{equation}
\rho_{NESS}=\Omega_{x}R\rho_{NESS}R\Omega_{x}%
\end{equation}
which explains why also $J_{\alpha=0,A=1}^{E}=0$. In fact, it can be easily
checked that under this symmetry the energy current operator changes sign,
$R\Omega_{x}\hat{J}^{E}R\Omega_{x}=-\hat{J}^{E}$.

Various anomalities in the steady currents are often visible at the level of
steady density profiles: e.g. ballistic current is usually accompanied by
magnetization profiles which are flat in the bulk. One might wonder if the
density profiles for our case, corresponding to the current anomalies at
$\alpha=0,\alpha=1$ are special. For the point $\alpha=0$, the exact $y$- and
$z$- magnetization profiles are trivial and flat, $\langle\sigma_{n}%
^{y}\rangle=$ $\langle\sigma_{n}^{z}\rangle=0$ for all $n$, a constraint,
imposed by the symmetry (\ref{SymmetryAlpha0}), while the $x$- profile
smoothly interpolates between the left and right boundary. On the other hand,
for $\alpha=1$ we do not find any particularity in the magnetization profiles,
which rather smoothly interpolate between the boundary values (even though
this can be a finite-size effect), data not shown.

\section{Conclusions}

\label{sec::Conclusions}

Transport properties of quantum systems can exhibit unexpected features if the
nonequilibrium steady state has to obey certain symmetry properties. Various
examples have been discussed in this paper for models of interacting systems
of qubits, subject to local pumping mechanisms from specific Lindblad
operators. We have first introduced different classes of these operators as
spin-targeting and dephasing ones. Then, we have discussed the kind of
symmetries they impose to the Lindblad master equation and to the
corresponding nonequilibrium steady state. The important role played by parity
symmetry selection rules has been illustrated for a general Hamiltonian model.
These considerations have been also specialized to the XXZ spin chain model.
We have shown that spin and energy currents can be suitably regulated by
acting on the symmetries of the NESS through the parameters of the Lindblad
operators. In particular, we find that both currents can vanish, even in the
presence of finite applied gradients.

We have to point out that all the results reported in this manuscript rely on
the basic assumption of uniqueness of the steady state solution of the
Lindblad Master equation. Such a property applies to all the examples
considered in this paper. An explicit check of this property can be performed
by using the completeness criterion of the algebra generated by the
Hamiltonian and by the Lindblad operators \cite{EvansUniqueness}. Once the
uniqueness is established, the nonequilibrium steady state is invariant under
all the symmetries of the Lindblad master equation. In fact, any violation of
a symmetry results in the existence of at least a one--parameter family of
steady state solutions as a direct consequence of the linearity of the
equation (\ref{LME}).

In a general perspective we can affirm that the properties of the steady
states analyzed in this manuscript can be viewed as a first achievement in the
exploration of new interesting features of the quantum Master equation. In
Sec.\ref{sec::1D driven XXZ model} we have also shown an explicit example of
how the vanishing of a current signals the presence of additional symmetries.

\textbf{Acknowledgements. } V.P. thanks the Center for Quantum Technologies,
National University of Singapore, where this work was initiated, for the kind
hospitality, and the Dipartimento di Fisica e Astronomia, Universit\`a di
Firenze, for support through a FIRB initiative. R.L. would like to acknowledge
financial support from the Italian MIUR-PRIN project n. 20083C8XFZ.


\begin{thebibliography}{99}
\bibitem {RemarkPsymmetry}NESS for $XXZ$ model enjoys full rotational symmetry
around the $z$-axis, represented by unitary operator $U(\beta)=\exp(i\beta
\sum_{k}\sigma_{k}^{z})$ , $\rho_{NESS}=U(\beta)\rho_{NESS}U(-\beta)$. The
$P$- symmetry (\ref{ParityConservation}) is a particular reduction of the
$U(\beta)$ for $\beta=\pi/2$.
\bibitem {RemarkOmegaYsymmetry}Another equivalent symmetry $\rho
_{NESS}=(\sigma^{y})^{\otimes_{N}}R\rho_{NESS}R(\sigma^{y})^{\otimes_{N}}$ is
obtained as a product of (\ref{OmegaXsymmetry}) and (\ref{ParityConservation}).
\end{thebibliography}
\end{document}